# An acoustic invisible gateway

Yi-Fan Zhu[1,2], Xin-Ye Zou[1,2], Bin Liang[1,2*], Wei-Wei Kan[1,2], Jun Yang[3], and Jian-Chun Cheng[1,2,3*]

[1]*Key Laboratory of Modern Acoustics, MOE, Institute of Acoustics, Department of Physics, Nanjing University, Nanjing* 210093*, P. R. China*

[2]*Collaborative Innovation Center of Advanced Microstructures, Nanjing University, Nanjing* 210093*, P. R. China*

[3]*Key Laboratory of Noise and Vibration Research, Institute of Acoustics, Chinese Academy of Sciences, Beijing 100190, P. R. China*

[*] To whom correspondence should be addressed. Emails: jccheng@nju.edu.cn (J. C. C.) and liangbin@nju.edu.cn (B. L.)

**Abstract**

The recently-emerged concept of "invisible gateway" with the extraordinary capability to block the waves but allow the passage of other entities has attracted great attentions due to the general interests in illusion devices. However, the possibility to realize such a fascinating phenomenon for acoustic waves has not yet been explored, which should be of paramount significance for acoustical applications but would necessarily involve experimental difficulty. Here we design and experimentally demonstrate an acoustic invisible gateway (AIG) capable of concealing a channel under the detection of sound. Instead of "restoring" a whole block of background medium by using transformation acoustics that inevitably requires complementary or restoring media with extreme parameters, we propose an inherently distinct methodology that only aims at engineering the surface impedance at the "gate" to mimic a rigid "wall" and can be conveniently implemented by decorating meta-structures behind the channel. Such a simple yet effective design of AIG also enables analyzing the physics behind this extraordinary phenomenon in an analytical manner, which agrees quite well with the numerical and experimental results. Furthermore, our scheme offers possibility to manipulate the symmetry of AIGs freely, which is proven both theoretically and experimentally by demonstrating two distinctive examples of one-way and two-way AIGs. We anticipate the realization of AIGs will open new avenues to illusion acoustics with potential applications in acoustic measurement and acoustic device, and provide inspirations for similar researches in other fields.

Illusion effects, which change the appearance of an object into that of another one of one's choice, have long fascinated people and inspired myths, novels, and movies. Recently, the pioneering works in coordinate transformation [1-4] and the concept of complementary media [5] have turned such imagines into scientific realities by enabling the design of conceptual illusion devices, which transform the scattered waves of an original object into that of another one chosen as illusion [6-12]. One such conceptual device is a gateway that is an open channel appearing to be blocked for waves of particular frequencies, which mimics a "hidden portal" mentioned in fiction [13] and has attracted great public interest [9-11]. Such an "invisible gateway" has been demonstrated for electromagnetic waves experimentally by Chan and coworkers by using a transmission-line medium which mimics a double-negative medium [11]. As another important classical wave, acoustic wave plays a crucial role in various applications ranging from noise reduction to acoustic detection to ultrasonic therapy. It should therefore be intriguing to realize a similar conceptual device with the extraordinary capability to conceal an open entrance to acoustic detections, which may also have great application potentials in many scenarios, e.g., underwater environments where acoustic wave is the sole possibility to reveal the existence of an object, or some important situations like acoustic measurements where one may desire to prevent an opening from affecting the environmental acoustic field while maintaining its ventilation property. A straightforward analogy between the optical and acoustic invisible gateways, however, is hardly feasible due to the inherent distinction between the electromagnetic and acoustic waves. Unlike negative electromagnetic parameters that can be found in natural materials (e.g., silver) [14], negative acoustic parameters have to be implemented by metamaterials and are usually accompanied by viscosity effect [15-19]. Considering the inherent difficulty in precise control of negative acoustic parameters required by coordinate transformation, it would be hard to analogously translate the concept of invisible gateway from electromagnetic waves to acoustic waves. So far the experimental realization of such a scientifically fascinating phenomenon of profound practical importance for acoustic waves still remains challenging.

In this article, we have reported a theoretical, numerical and experimental work on the design of the first "acoustic invisible gateway" (AIG) capable of concealing a channel acoustically. Here we focus our attention on the rational design of surface impedance, instead of employing transformation-based schemes generally used in designs of optical invisible gateway. In the context of coordinate transformation theory, the blockage of electromagnetic waves by an open channel has to be achieved by optically "restoring" a piece of space filling the channel using its restoring medium with precisely-controlled double-negative parameters [9-11]. In our design, contrarily, we directly manipulate the surface impedance at the open "gate" to mimic that of a rigid "wall", which is a more straightforward and fundamental manner to realize the desired effect of invisibility gateway. Thanks to such an inherently distinct scheme of "surface" design rather than "volumn" design, an AIG can be implemented simply by decorating meta-structure at the back of wall to adjust the effective acoustical impedance at the front surface, without exerting demanding requirements on the properties of the whole block of media filling the gateway. This also enables an analytical analysis on the equivalent impedance based on transmission-line theory that helps to predict the transmittance and phase shift. Numerical simulations and experimental measurements, agreeing quite well with the theoretical predictions, demonstrate the performance of the designed AIG device. Furthermore, our design enables freely manipulating the symmetry of the AIG effect, which has also been verified both numerically and experimentally by demonstrating two distinct examples of one-way and two-way AIGs. With the capability of rendering an open gateway undetectable acoustically and flexibility of controlling the symmetry of such invisibility effect, our design of AIG should have deep implication in acoustic research and all acoustic-based applications. Our simple yet effective scheme of yielding AIG effect by directly tailoring the surface impedance may also offer inspiration for design and implementation of similar invisible-gateway-like devices in other fields like electromagnetics.

**Results**

**Theoretical design of an AIG.** First we will explain our design idea of an AIG. In transformation-based schemes, one needs to "close" the gateway acoustically by adding restoring media, which results in a demanding condition on its implementation that requires precise control of double-negative parameters [15-19]. Here we propose an essentially distinct approach that uses meta-structures behind the wall, as shown in Fig. 1(a), to conceal the channel without changing the original straight shape of the channel or partially blocking this passage for other entities. As shown in Fig. 1(b), we assume the $x$-components of incidence acoustic pressure and reflected acoustic pressure are $p_i e^{-ikx}$ and $\beta p_i e^{i(kx-\phi)}$ ( $0 \leq \beta \leq 1$ ), respectively, where $\beta$ represents the ratio between the acoustic pressures of the reflected and incident waves, and $\phi$ indicates the phase difference. In order to conceal a channel, when the acoustic wave impinges on the channel, the reflected wave should be identical with the one caused by an intact wall, i.e., a flat wall without any opening. This requires $\beta = 1$ and $\phi = 0$, which means that the reflected wave undergoes a total reflection and has unchanged phase. These two values will be influenced by the effective characteristic acoustic impedance at the entrance, which can be expressed as

$$Z_{\text{eff0}} = \frac{p_t}{v_t} = Z_0 \frac{p_i e^{-ikx} + \beta p_i e^{i(kx-\phi)}}{p_i e^{-ikx} - \beta p_i e^{i(kx-\phi)}} = Z_0 \frac{1 + \beta e^{i\phi}}{1 - \beta e^{i\phi}} , \qquad (1)$$

where $p_t$ and $v_t$ are total acoustic pressure and particle velocity, respectively, $Z_0 = \rho_0 c_0$ is the characteristic impedance of the background medium that is chosen as air here, with $\rho_0 = 1.21 \text{kg}/\text{m}^2$ and $c_0 = 343 \text{m/s}$ being the mass density and sound velocity respectively. Equation (1) gives the dependence of acoustic impedance on the reflected acoustic pressure ratio and phase difference. Observing Eq. (1), $\beta = 1$ and $\phi = 0$ requires $Z_{\text{eff}0}$ to be equal to the acoustical impedance of a rigid wall that has an infinite real part and a zero imaginary part, viz., $R \gg Z_0$ and $X = 0$ with $R = \text{Re}(Z_{\text{eff}0})$ and $X = \text{Im}(Z_{\text{eff}0})$ respectively. Under such circumstance, one can deduce from Huygens' principle [20] that the whole reflected field will not be affected by the presence of an open channel and will be identical with an intact wall, since the *y*-component is independent of the effective impedance interface. On the other hand, an AIG also needs to prevent the observer from "hearing" acoustic wave at the other side of wall through the channel, which can be spontaneously be satisfied as long as one has $\beta = 1$ according to the reciprocity principle in such a linear system.

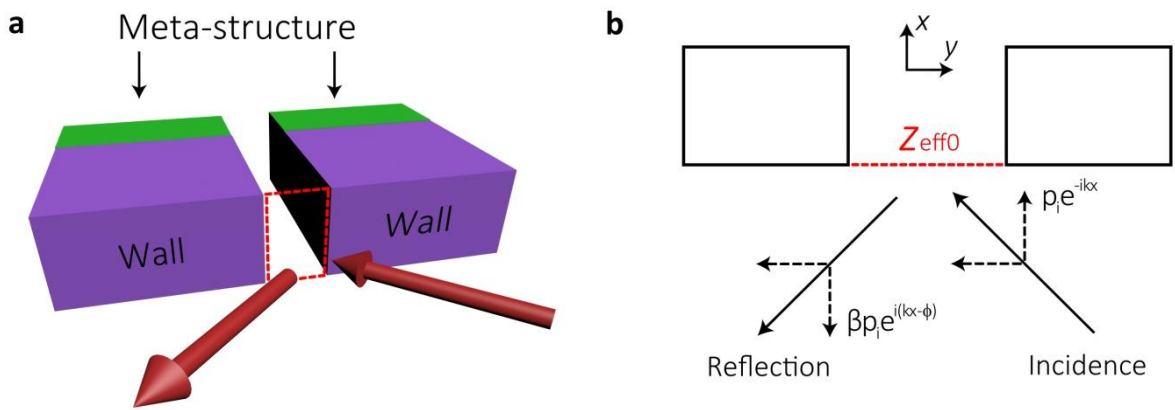

**Figure 1 | Schematic illustration of AIG.** (**a**) Schematic of an AIG that consists of an open channel and well-designed meta-structures at the back of walls. The incident sound will

reflect at the gate of channel, as if there is no channel at all. (**b**) The effective impedance at the gate, $Z_{eff0}$, can be modulated by the designed meta-structures, and will determine the reflected acoustic wave.

**Implementation of AIG by meta-structures.** Next we show how a practical acoustic meta-structure decorated at the back of wall can yield the desired effective impedance at the gate while maintaining a straight open channel. Here we implement the meta-structures by a periodic array of Helmholtz resonators (HRs), as Fig. 2(a) shows, which has not altered the basic shape of the channel with a width of $D = 5$cm. The acoustic impedance of HR can be expressed as $Z_H = i(1/\omega C_H - \omega M_H)$, where $\omega$ is angular frequency, $C_H$ and $M_H$ are acoustic capacitance and acoustic mass of the HR, respectively, and $d$ is the period of HR arrays. The propagation of the incident wave in this system can be analyzed by using acoustic transmission-line method [21-23]. As shown in Fig. 2(b), the radiant energy at the export is decided by the local impedance $Z_{eff1}$ that will be affected by the meta-structure transmission-lineas the equivalent circuit model shows. In general, since the HRs are open to free spaces, acoustic wave may radiate out at the interface of the meta-structure, which can be expressed by a shunt-wound impedance with $Z_0$ in transmission-line circuit as the upper part in Fig. 2(b) shows. This seems to violate the above-mentioned requirement for achieving total reflection. However, when the effective impedance at every forks of the circuit is much lower than $Z_0$, the branch of $Z_0$ can be seen as open circuit and the circuit can be simplified into the bottom one shown in Fig. 3(b), just similar to the case of HRs in a waveguide structure [23-24]. Such a special condition can be achieved under particular resonances, by selecting appropriate values for $Z_H$ and $d$, which is referred to as "bounding points" where the wave is bounded at the surface instead of radiating out. When the incident frequency $f_0$ is chosen close to the resonant frequency of an individual HR, which can be easily calculated to be $f_r = \sqrt{1/M_H C_H}/2\pi$, the impedance of HRs will become

$$Z_H = i\left[\frac{C_H}{\omega_r + \Delta\omega} - (\omega_r + \Delta\omega)M_H\right] = -2i\Delta\omega M_H, \tag{2}$$

where $\omega_0 = \omega_r + \Delta\omega$ is the incident angular frequency and $\omega_r = \sqrt{1/M_H C_H}$. Assuming the number of the HRs is $N$, the impedance at $N$th port is $Z_N = Z_H Z_0/(Z_H + Z_0)$ which is parallelly connecting $Z_H$ and $Z_0$. Then, the transferred impedance at $Z_{N-1}$ can be achieved by applying acoustic impedance transfer formula [21]

$$Z_{l'} = Z_0 \times \frac{Z_l + jZ_0 \tan(kl)}{Z_0 + jZ_l \tan(kl)}, \tag{3}$$

where $l$ is the propagation distance and equal to period $d$ in our model. If transfer impedance is much lower than $Z_0$, the branch of $Z_0$ will be negligible and no wave will radiate to free space. The impedance at $N-1$th port can be calculated by parallelly connecting this transferred impedance and $Z_H$. This process should be repeated for $N$ times. Thus, the bounding point requires this iteration process to be convergent and the impedances at every fork to be nearly zero, at which the equivalent circuit can be simplified. In this way, $Z_{\text{eff1}}$ can be obtained, and the reflected power ratio at the export can then be expressed as [21]

$$r_{\mathrm{I}} = \frac{(R-Z_0)^2 + X}{(R+Z_0)^2 + X}. \tag{4}$$

When the load at the end of channel is a rigid object ($R \gg Z_0$), total reflection ($r_{\mathrm{I}} = 1$) will occur. However, Eq. (4) indicates that there is another total reflection condition, that is $R \to 0$. Apparently, this impedance value, which represents a medium much more soft than air, is absent in natural materials. But in the current meta-structure, such a nearly zero value of $R$ can be obtained by utilizing resonant effect with proper structural parameters. Figure 2(c) gives the real part of $Z_{\mathrm{eff1}}$ value versus period $d$ and incident frequency $f_0$. The desired near-zero values only appear at the frequency slightly lower than $f_{\mathrm{r}}$ with proper values of $d$. This is due to the fact that when $f_0 > f_{\mathrm{r}}$, we get from Eq. (2) that $X_{\mathrm{H}} < 0$, which makes the above iteration process always divergent and unable to get a sufficiently low $R$. The convergence can only be satisfied when $X_{\mathrm{H}} > 0$. Therefore, by introducing both acoustical capacitance and mass element into a channel structure with periodic decorations [25-26], the effective resistance at the export can be modulated to be nearly zero, and the wave could totally reflect back instead of leaking from the opening of the channel. The proposed structure also provides greater flexibility to obtain desired impedance for different frequencies. In experiment, we choose incident frequency as $f_0 = 2920 \mathrm{Hz}$ with $d = 2.5 \mathrm{cm}$ and $Z_{\mathrm{eff1}} = 240i + 26 kg \cdot m^{-2} \cdot s^{-1}$ which has a sufficiently low real part, compared with the impedance of air $Z_0 = 415 kg \cdot m^{-2} \cdot s^{-1}$.

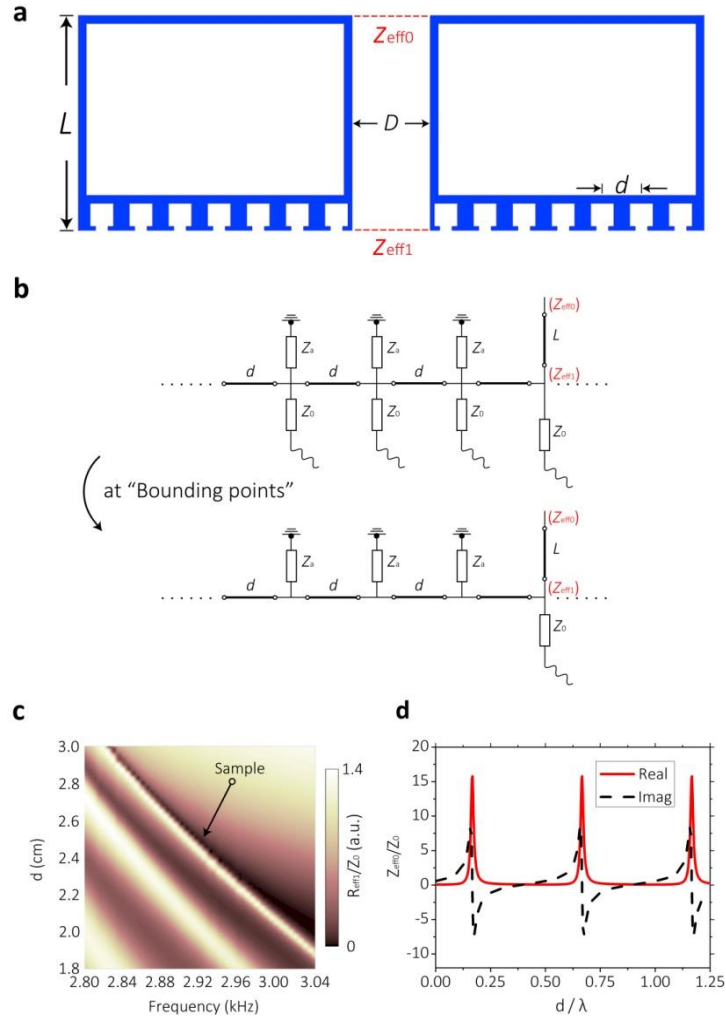

**Figure 2 | Design of extraordinary effective acoustic impedances.** (**a**) 2-D schematic of the designed AIG implemented by meta-structures. The channel width is set as $D = 5\text{cm}$. (**b**) Impedance analysis by acoustic transmission-line method. At bounding points, the upper equivalent circuit can be simplified into the bottom one. (**c**) The effective acoustic impedance at the export of the channel $Z_{\text{eff1}}$ versus period $d$ and frequency $f_0$. The period of the sample is set as $d = 2.5\text{cm}$ for $f_0 = 2920\text{Hz}$ with $Z_{\text{eff1}} = 240i + 26 kg \cdot m^{-2} \cdot s^{-1}$ which has a sufficiently low real part as compared with $Z_0 = 415 kg \cdot m^{-2} \cdot s^{-1}$, the impedance of air. (**d**) The effective acoustic impedance at the orifice $Z_{\text{eff0}}$ versus $L$. The case of $R_{\text{eff0}} \gg Z_0$ and $X_{\text{eff0}} \to 0$ appears at $\lambda = L/(0.018 + n/2) \quad (n = 0, 1, 2...)$.

The above designs of $Z_{eff1}$ are sufficient to achieve total reflection at the export of the channel. To makes the reflected phase matched at the entrance of the channel, the effective impedance $Z_{eff0}$ at the orifice should be consistent with that of intact wall as Eq. (1) indicates. Thus, we apply impedance transfer formula in Eq. (3) again and obtain the relationship between $Z_{eff0}$ and $L$ in Fig. 2(d). In experiment, the length is $L=13.85$cm and we get a desired $Z_{eff0}$, which is purely resistive and much larger than $Z_0$, and reflected acoustic wave could be expected to be identical with that caused by an intact wall.

**Numerical and experimental demonstrations of AIG**. The comparisons between numerical and experimental results of AIG are shown in Fig. 3. Figure 3(a) shows the comparison between the simulated acoustic pressure distributions for 'intact wall', 'AIG' and 'channel' at 2920Hz. The parameter of reflected field difference (RFD) is defined as $\text{RFD}=|p(x,y)-p_{wall}(x,y)|$, where $p(x,y)$ is the acoustic pressure at $(x,y)$ and $p_{wall}(x,y)$ represents the one caused by an intact wall. The bounding effect can be observed in AIG case, in which only evanescent wave appears at the interface of the meta-structure and no wave radiates out. The corresponding measured RFDs in Fig. 3(a) show that only the reflected field distributions for AIG are identical with an intact wall. We have also measured the transmitted fields through the AIG and the channel in Fig. 3(b). The numerical and experimental results for AIG reveal that the transmitted fields are nearly zero and the wave is unable to pass through the AIG. In comparison, the RFD and transmitted field for a channel in Figs. 3(a) and 3(b) are obviously seen. The comparison between AIG and channel demonstrates the invisibility effect in the well-designed AIG. The photograph of the sample and the experiment setup are shown in Fig. 3(c), and the details are displayed in "Methods" section. Figure 3(d) gives the simulated and measured reflected phase and amplitude changes for three particular cases. For an ordinary channel, the reflected amplitude is much lower than the incident one. We have also obtained another case that the channel is blocked at the end and can thus be regarded as a "defect" for comparison, which give rise to total reflection but phase mismatched. The AIG structure has both right reflected phase and amplitude as the theoretical and the measured values of AIG in Fig. 3(d) show.

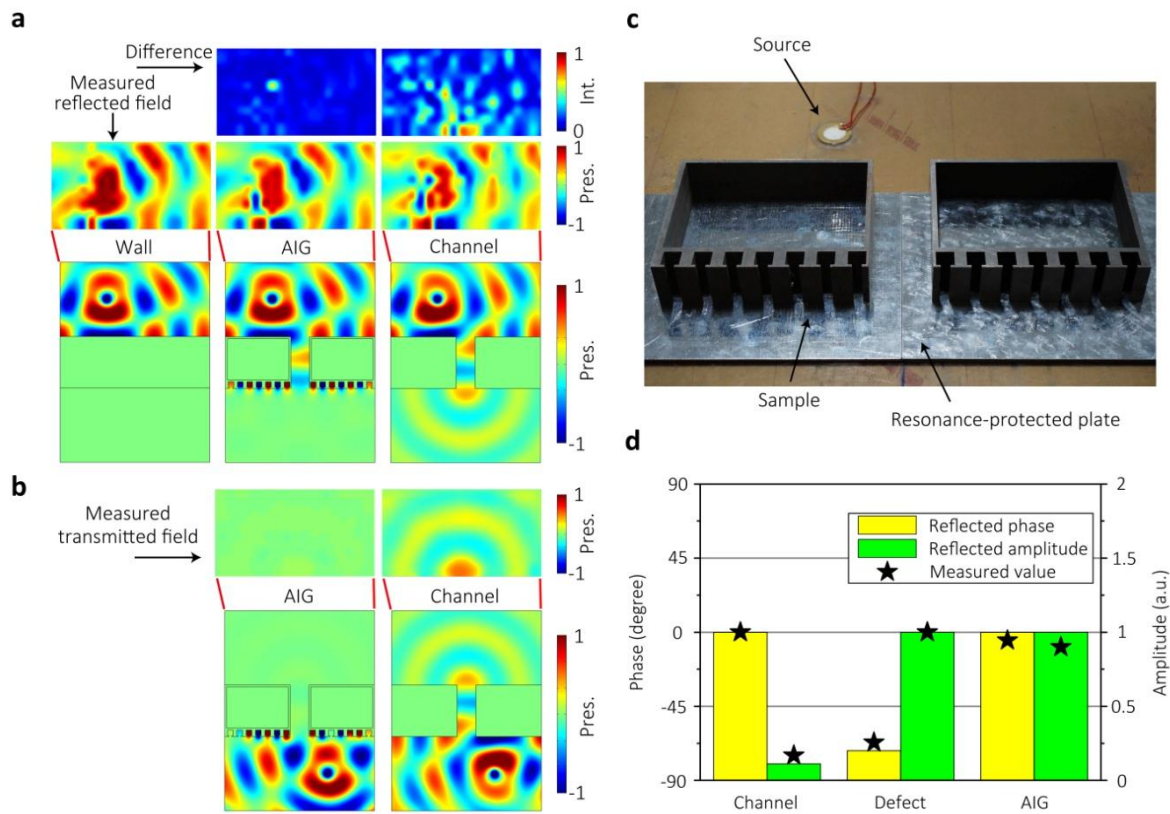

**Figure 3 | Numerical and experimental demonstrations of AIG**. (**a**) The simulated and measured acoustic pressure fields for 'intact wall', 'AIG', 'channel' cases, and the

corresponding RFDs at 2920Hz. The RFDs are calculated to demonstrate the invisibility effect. (**b**) The simulated and measured transmitted fields for AIG and a channel, respectively. (**c**) The photograph of the AIG sample and the experimental setup. The top plate of the waveguide (not shown in the diagram) is located 4.6cm above the bottom plate. Steel plates with a thickness of 0.3cm are placed at both the bottom and the top of the samples as a resonance-protector to prevent the resonant peaks from slightly shifting that would occur if resonant cavity is directly connected with the parallel waveguide plate made of plexiglass. (**d**) The simulated and measured reflected phase and amplitude for these particular conditions.

**One-way and two-way invisibilities.** Our scheme also offers possibility to manipulate the symmetry of AIGs freely. The original structure in Fig. 2 is designed to have only one-way invisible functionality due to that at the back of the wall, the HR array and undesired acoustic impedance at the orifice will make the reflected fields quite different with an intact wall, making the back side can be detected by sound. This one-way invisible characteristic may be useful for some occasions where the "front" and "back" sides of the AIG need to be distinguishable such that the observer behind the AIG could be aware where the "gate" is. On the other hand, our proposed scheme also has the potential to design two-way acoustic invisible gateway (TAIG) which could not be detected from the both sides of the wall. Figure 4(a) displays the schematic of TAIG with a sandwich structure. The HRs array is now placed in a branch channel and we set the width of the branch to be $l = 2.5$cm as marked in Fig. 4(a). The corresponding impedance analysis is shown in Fig. 4(b). By setting HRs in such a closed space, the equivalent circuit here is always a simplified one, which is similar to the one at bounding points for a one-way AIG. Figure 4(c) gives the relationship between the effective impedance of $Z_{\text{eff1}}$ and period $d$. The near-zero value appears at $d = 2.45\text{cm}$ and $f_0 = 2920\text{Hz}$, which is close to the parameter of one-way AIG ($d = 2.5\text{cm}$ and $f_0 = 2920\text{Hz}$). This agreement in turn verifies that at bounding point, the wave is confined in a closed space. The slightly shift is due to the fact that the end of the tube is now closed, and can be expressed as ground in equivalent circuit, which produces a slightly-changed initial value. Theoretically, the circuit of the channel part should be symmetric to obtain the same extraordinary impedance at the orifices of front and back sides, viz., the length $L$ should be same for both sides in Fig. 4(b). In practice, however, since the width of $l = 2.5\text{cm}$ is not deep subwavelength, the lengths of channel in Fig. 4(a) are properly-tuned and we set them as $L_1 = 16\text{cm}$ and $L_2 = 14.8\text{cm}$. In order to demonstrate the two-way invisible effects, we set two point sources at the front and back sides of the channel respectively to mimic a practical condition that sound are generated at the both sides of the channel. The numerical acoustic field with two point sources incidence is shown in Fig. 4(d), and we have measured the acoustic fields at the both sides of the TAIG. Comparing the simulated and measured RFDs in Figs. 4(e) and 4(f), the invisible phenomena are observed at the both sides, verifying the effectiveness of the TAIG. Figures 4(g) gives the photograph of TAIG samples and the details are in the "Methods" section. In the two-way structure, it is also convenient to experimentally obtain the transmitted ratio. The transmitted pressure ratios for two directions should be consistent on the basis of reciprocity principle and the measured values at different frequencies are displayed in Fig. 4(h). The measured results are in good agreement with the simulated ones, suggesting that the acoustic wave is unable to pass through the AIG structure at $f_0 = 2920\text{Hz}$, compared with an ordinary channel.

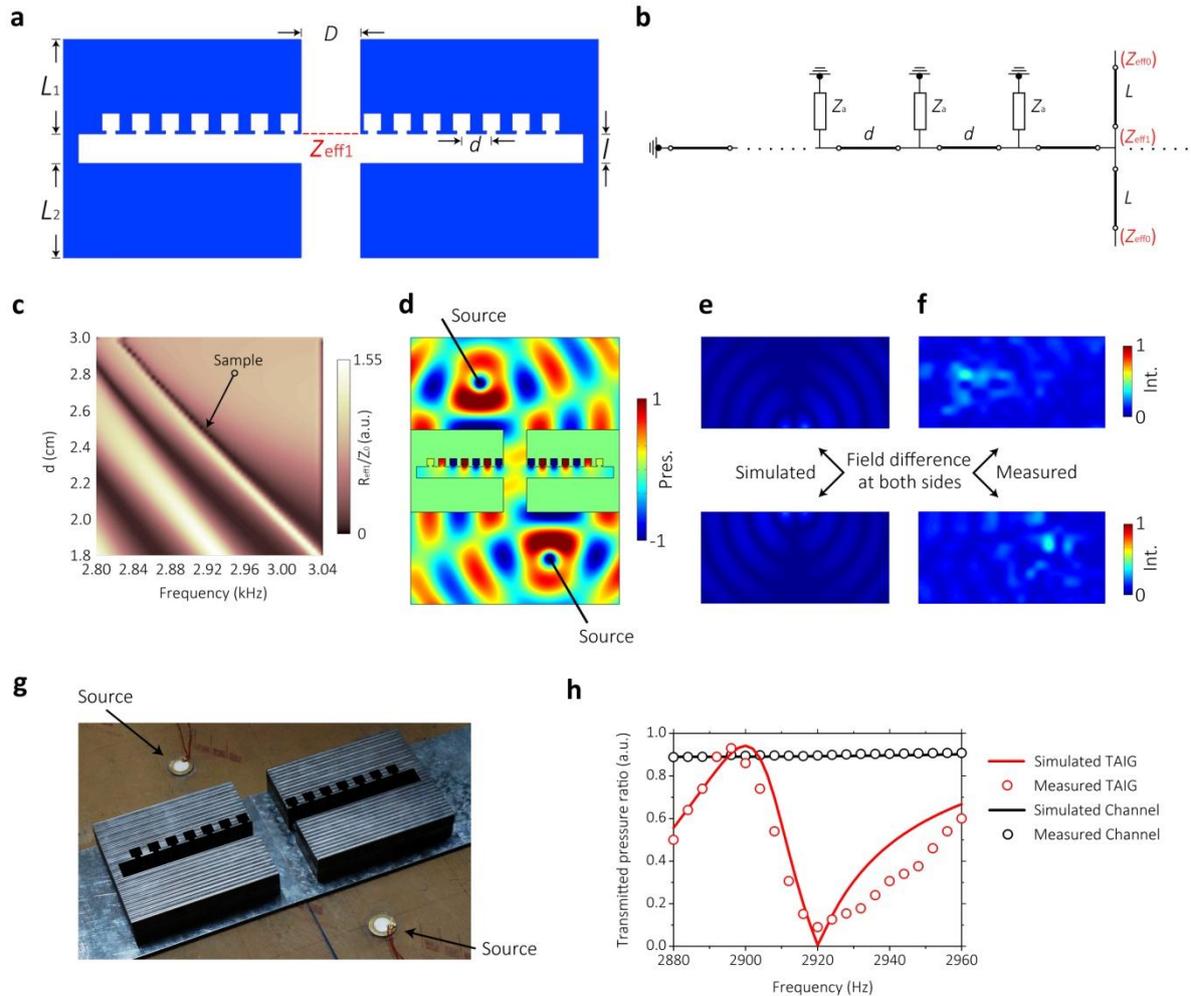

**Figure 4 | Numerical and experimental demonstrations of TAIG**. (**a**) 2-D schematic of TAIG. The channel width keeps $D = 5\text{cm}$. (**b**) Impedance analysis by acoustic transmission-line method. (**c**) The effective acoustic impedance at the export of the channel $Z_{\text{eff1}}$ versus period $d$ and frequency $f_0$. The period is set as $d = 2.45\text{cm}$ for $f_0 = 2920\text{Hz}$ with $Z_{\text{eff1}} = 194i + 0.01 kg \cdot m^{-2} \cdot s^{-1}$. (**d**) The simulated acoustic fields with two

point sources incidence. (**e**) The simulated front and back RFDs. (**f**) The measured front and back RFDs. (**g**) The photograph of the TAIG sample and the experimental setup. The top plate of the waveguide (not shown in the diagram) is located 4.6cm above the bottom plate. (**h**) The numerical and experimental transmitted pressure ratios for TAIG and an ordinary channel with plane wave incidence.

**Discussion**

We have proposed a new way to realize AIG by designing effective acoustic impedance at the entrance of the channel. The invisible effects are demonstrated by numerically calculating and experimentally measuring the RFDs and transmitted fields. The proposed HR arrays structure at the back of the wall shows the possibility to manipulate the effective acoustic impedance at the orifice freely. The structural parameter $f_r$ and $d$ can be adjusted to achieve desired working frequency $f_0$ and the channel length $L$ to satisfy the practical need. One-way and two-way functionalities are also optional and by choosing the parameters of TAIG, the values of $f_0$, $L_1$, and $L_2$ are also tunable. Our finding may pave the way for design of illusion-acoustics device and have applications in various occasions such as acoustic measurement and architectural acoustics.

**Methods**

**Experimental setup and sample fabrication.** Our experiment was accomplished in the lab-made parallel waveguide plate system [27-28] a tunable height. The parallel waveguide plate is made of plexiglass (polymethyl methacrylate) with $c_{glass} = 2700$m/s and $\rho_{glass} = 1200$kg/m$^2$. The samples are made of steel with $c_{steel} = 5200$m/s and density $\rho_{steel} = 7850$kg/m$^2$ and processed by WEDM (Wire Electrical Discharge Machining) technology. The samples in Figs. 3(c) and 4(g) have 3-D sizes of $17.5$cm$\times 13.85$cm$\times 4$cm and $20$cm$\times 17.9$cm$\times 4$cm, respectively. In measurement, we place the sample between two steel plates with a thickness of $0.3$cm to prevent the slight shift of resonant peaks that may occur if the resonant cavity is directly connected with the parallel waveguide plate made of plexiglass. Such a slight shift is due to that our proposed structure is resonance-sensitive and the parameter difference between plexiglass and air is relatively low. Thus, the height between top and bottom plate in Figs. 3(c) and 4(g) is $4.6$cm. Two 0.25-inch-diameter Bruel & Kjær type-4961 microphones are used to detect the acoustic fields. One microphone is used to scan the measuring region with a size of $40$cm$\times 20$cm and the other is fixed near the loudspeaker as reference microphone to obtain both the amplitude and phase information.

**Numerical simulations.** The numerical simulation results are conducted by the Finite Element Method based on commercial software COMSOL Multiphysics$^{TM}$ 4.3. The simulated material for has the same parameters with the samples. The surrounding medium is air.